\documentclass[aps,prd,twocolumn,superscriptaddress,showpacs,amsmath,amssymb,nofootinbib, preprintnumbers]{revtex4-1}
 \usepackage{amsmath}
\usepackage{graphicx}
\usepackage{epstopdf}
\usepackage{float}
\usepackage{color}
\usepackage[toc,page]{appendix}
\usepackage[usenames,dvipsnames]{xcolor}
\usepackage[normalem]{ulem}

\newcommand{\be}{\begin{equation}}
\newcommand{\ee}{\end{equation}}
\newcommand{\ba}{\begin{eqnarray}}
\newcommand{\ea}{\end{eqnarray}}

\newcommand{\beq}{\begin{equation}}
\newcommand{\eeq}{\end{equation}}
\newcommand{\beqa}{\begin{eqnarray}}
\newcommand{\eeqa}{\end{eqnarray}}

\begin{document}

\title{Misner Gravitational Charges and Variable String Strengths}

\author{Alvaro Ballon Bordo}
\email{aballonbordo@perimeterinstitute.ca}
\affiliation{Perimeter Institute, 31 Caroline Street North, Waterloo, ON, N2L 2Y5, Canada}
\affiliation{Department of Physics and Astronomy, University of Waterloo,
Waterloo, Ontario, Canada, N2L 3G1}

\author{Finnian Gray}
\email{fgray@perimeterinstitute.ca}
\affiliation{Perimeter Institute, 31 Caroline Street North, Waterloo, ON, N2L 2Y5, Canada}
\affiliation{Department of Physics and Astronomy, University of Waterloo,
Waterloo, Ontario, Canada, N2L 3G1}

\author{Robie A. Hennigar}
\email{rhennigar@mun.ca}
\affiliation{Department of Mathematics and Statistics, Memorial University of Newfoundland, St. John’s, Newfoundland and Labrador, A1C 5S7, Canada}

\author{David Kubiz\v n\'ak}
\email{dkubiznak@perimeterinstitute.ca}
\affiliation{Perimeter Institute, 31 Caroline Street North, Waterloo, ON, N2L 2Y5, Canada}
\affiliation{Department of Physics and Astronomy, University of Waterloo,
Waterloo, Ontario, Canada, N2L 3G1}

\date{May 30, 2019}

\begin{abstract}
As shown recently \cite{Kubiznak:2019yiu},  consistent thermodynamics of the Lorentzian Taub--NUT solutions with Misner strings present can be formulated provided a new pair of conjugate quantities (related to the NUT parameter) $\psi-N$ is introduced. In \cite{Kubiznak:2019yiu} this pair was calculated from the Euclidean action but no geometrical interpretation for the new quantities was provided. In this paper we propose that the potential $\psi$ should be identified with the surface gravity of the Misner string and the conjugate Misner charge $N$ can be obtained by a Komar-type integration over the tubes surrounding
the string singularities. We show that similar tube contributions also modify the Komar formula for the thermodynamic volume. To render the integrals finite we  employ the method of Killing co-potentials. By construction the new charges then satisfy the Smarr relation.
Equipped with these geometrical notions, we generalize the first law for the {(possibly charged)} Taub--NUT spacetimes to account for asymmetric distributions of Misner strings and their potential variable strengths.
\end{abstract}

\maketitle

\section{Introduction}

The Lorentzian Taub--NUT spacetime \cite{taub1951empty, newman1963empty} is full of surprises.
Besides the Schwarzschild-like horizon, it features  the unprecedented
rotating string like singularities (Misner strings) on the axis and the associated regions with closed timelike curves (CTCs) in their vicinity.

Over the years, two proposals for the physical interpretation of the Taub-NUT spacetime have emerged.
Misner \cite{misner1963flatter}, studying the group of motions generated by the Killing vectors on an $r = const.$ hypersurface, argued that these sections of the spacetime are topologically three-spheres and that the time coordinate should be periodically identified to ensure geometric regularity. This, Misner found, was sufficient to remove the string singularities along the axis, but comes at the price of introducing CTCs everywhere in the spacetime.
An alternate interpretation was offered by Bonnor \cite{bonnor1969new}, who argued that the Misner string singularities---and associated CTC regions---should be replaced by a corresponding matter source, some form of rotating topological defect. This provides a physical interpretation for the `lack of regularity' in the unidentified spacetime, somewhat analogous to spacetimes containing cosmic strings~\cite{Israel:1976vc}.

However, it has been recently understood that the Taub-NUT spacetime with Misner strings present is less pathological than previously thought, and their complete removal may be unnecessary~\cite{Clement:2015cxa, Clement:2015aka, Clement:2016mll}. The Misner strings are transparent for geodesics, rendering the spacetime geodesically complete, and despite the occurrence of CTC regions, the spacetime is (at least for sufficiently small string strengths) free of causal pathologies for geodesic observers  \cite{miller1971taub, Clement:2015cxa, Clement:2015aka, Clement:2016mll}. In this paper we thus consider the whole spacetime with Misner strings present as `physically relevant' and study its thermodynamic properties.

The thermodynamics of  the Lorentzian Taub--NUT solutions with Misner strings present has been recently formulated in \cite{Kubiznak:2019yiu} (see also \cite{Ballon:2019uha} for the extension to the $U(1)$ charged case).
The key step was to introduce a new pair of conjugate quantities $\psi-N$, allowing for the independent variations of the NUT parameter in the first law. For the uncharged Taub--NUT-AdS solution such a new first law then takes the following form:
\be\label{first1}
\delta M=T\delta S+\psi \delta N+V\delta P\,.
\ee
Here, $M$ stands for the mass of the solution, $T$ is the temperature of the `black hole horizon', $S$ is the corresponding entropy given by the Bekenstein--Hawking area law,\footnote{In this paper we do not pursue the second proposal put forward in \cite{Kubiznak:2019yiu}, where the entropy is identified with the `Noether charge' \cite{Garfinkle:2000ms}, which leads to some pathological features.}   $P$ is the cosmological constant pressure and $V$ is the conjugate thermodynamic volume,
\be\label{P}
P=-\frac{\Lambda}{8\pi}\,,\quad V=\Bigl(\frac{\partial M}{\partial P}\Bigr)_{S,N}\,.
\ee
However, in \cite{Kubiznak:2019yiu} no physical interpretation or geometrical meaning were given to the two new quantities $\psi$ and $N$. It is a purpose of this paper to amend this situation.

As already noted in \cite{Carlip:1999cy}, at the positions
 of Misner strings there are Killing horizons.\footnote{Let us stress here that the corresponding Killing generators are different on the north and south pole axes and are different from the generator of the black hole horizon.}  In what follows we propose to identify the corresponding surface gravity with the potential $\psi$. In fact, for an asymmetric distribution of Misner strings the north pole and south pole surface gravities are different and yield thus two `independent' potentials $\psi_\pm$. The corresponding conjugate quantities $N_\pm$, we call them the gravitational Misner charges, can
be obtained by a Komar-type integration over the tubes surrounding the Misner string singularities.
Such an integral is finite for the asymptotically flat Taub--NUTs but diverges in the AdS case where it has to be `renormalized'. We show that similar to the black hole volume, this renormalization can be achieved with the help of Killing co-potentials \cite{Kastor:2009wy}.
The prescription for $N_\pm$ (see \eqref{Nint} below) originates from a careful re-derivation of the generalized Smarr relation, taking into account the new boundaries induced by the presence of Misner strings. Interestingly, the tube integrals also modify the `standard prescription' for the thermodynamic volume $V$, see \eqref{Vint} below.  Of course, by construction the resultant quantities obey the generalized Smarr relation
\be
M=2(TS+\psi_+N_++\psi_-N_--PV)\,,
\ee
and (with a careful choice of the Killing co-potential) also the extended first law
\be\label{first2}
\delta M=T\delta S+\psi_+\delta N_++\psi_- \delta N_-+V\delta P\,.
\ee
Such a first law not only accounts for possible asymmetric distribution of Misner strings, but also the `strength' of the overall Misner string can be independently varied. In this sense it is of full cohomogeneity and reduces to \eqref{first1} upon fixing the Misner string strengths.

Our paper is organized as follows. In the next section we review the basic facts about the Taub--NUT-AdS solution and discuss the Misner Killing horizons.
The thermodynamic argument of \cite{Kubiznak:2019yiu} leading to the introduction of the new thermodynamic conjugate pair $\psi-N$ is briefly summarized in Sec.~\ref{sec3}. Sec.~\ref{sec4} is devoted to the derivation of the generalized Smarr relation; novel Komar-like formulae for the Misner charges $N_\pm$ and the volume $V$ are presented. In Sec.~\ref{sec5} such formulae are applied to the Taub--NUT-AdS spacetime to derive the generalized first law for an arbitrary distribution  of Misner strings.
{Sec.~\ref{sec6} remarks on a potential physical interpretation of Misner charges.}
Summary and final discussion are presented in Sec.~\ref{sec7}.
{The appendix is devoted to the generalized first law for the dyonic Taub--NUT solutions. }

%%%%%%%%%%%%%%%%%%%%%%%%%%%%%%%%%%%%%%%%%%%%%%%%%%%%%%%%
%%%%%%%%%%%%%%%%%%%%%%%%%%%%%%%%%%%%%%%%%%%%%%%%%%%%%%%
\section{Geometry of Taub--NUT-AdS spacetime}\label{sec2}

The AdS Taub--NUT solution reads
\ba\label{j1}
ds^2&=&-f\bigl(dt+2n\cos\theta d\phi\bigr)^2+\frac{dr^2}{f}\nonumber\\
&&\qquad \quad +(r^2+n^2)(d\theta^2+\sin^2\!\theta d\phi^2)\,,\\
f&=&\frac{r^2-2mr-n^2}{r^2+n^2}- \frac{3n^4-6n^2r^2-r^4}{l^2(r^2+n^2)}\,,
\ea
where $n$ stands for the NUT charge, $m$ for the mass parameter, and $l$ for the AdS radius,
\be
\Lambda=-\frac{3}{l^2}\,.
\ee

The above metric describes a situation where the Misner strings are `symmetrically distributed': both the south $(\theta=\pi)$ and the north ($\theta=0)$ pole axis are `equally singular'. As we shall show in this paper, for formulating thermodynamics this is not required and an `arbitrary' distribution  of these defects,   or strength of Misner strings, can be considered. The corresponding metric can be obtained by  the following `large coordinate transformation' \cite{Clement:2015cxa}
\be
t\to t+{2s \phi}\,,
\ee
corresponding to `threading the spacetime with an overall Misner string' (changing effectively the strengths of north pole and south pole Misner strings),
upon which we obtain
\ba\label{metric2}
ds^2&=&-f\Bigl[dt+2(n\cos\theta +s) d\phi\Bigr]^2+\frac{dr^2}{f}\nonumber\\
&&\qquad \quad +(r^2+n^2)(d\theta^2+\sin^2\!\theta d\phi^2)\,.
\ea
We stress that the parameter $s$ is a new physical dimensionful parameter. Roughly speaking, whereas $n$ corresponds to the differential strength  between the two Misner strings, $s$ governs their (overall) absolute magnitude.\footnote{The situation is somewhat analogous to that of the C-metric, see \cite{Anabalon:2018qfv}, where the acceleration parameter $A$ corresponds to the difference between the two cosmic string tensions (causing the black hole to accelerate), and the conical deficit parameter $K$ tunes the (overall) absolute magnitude.    }
In particular, the choice of $s=0$ recovers the symmetric distribution of strings, while $s=\pm n$
makes the south (north) axis regular.\footnote{According to Misner \cite{misner1963flatter}, all such strings are unobservable provided the time is identified as $t\sim t+8\pi n$\,. In what follows we shall not impose this condition and consider the time coordinate non-compactified.}
As shown in \cite{Clement:2015cxa}, the (asymptotically flat) spacetime is geodesically complete for any value of $s$, but the requirement for the absence of closed timelike and null geodesics requires  $|s/n|\leq 1$.

There is a Killing horizon in the spacetime, located at the largest root $r_+$ of
$f(r_+)=0$ and generated by the Killing vector
\be
k=\partial_t\,.
\ee
We shall refer to this horizon as a black hole horizon. The associated temperature is
simply given by
\be\label{T}
T=\frac{f'}{4\pi}=\frac{1}{4\pi r_+}\Bigl(1+\frac{3(n^2+r_+^2)}{l^2}\Bigr)\,,
\ee
while its area reads
\be
\mbox{Area}=\pi(r_+^2+n^2)\,.
\ee

There are yet other Killing horizons present in the spacetime. Namely, when the Misner string is present, the north/south pole axis is a Killing horizon of the following Killing vector:
\be\label{kpm}
k_\pm=\partial_t- \frac{1}{2(s\pm n)} \partial_\phi\,.
\ee
Note that, contrary to $k$, these Killing vectors are not properly normalized at infinity and their norm is there $\theta$-dependent. The corresponding surface gravities can be calculated using the standard formula:
\be
\kappa^2=\frac{1}{4}\frac{\nabla^\mu L \nabla_\mu L}{L}\,,\quad L=-k^2\,,
\ee
which yields
\be
\kappa_\pm=\frac{1}{2(n\pm s)}\,.
\ee
In what follows we shall associate with these the following `Misner potentials':\footnote{{The normalization of the surface gravities of the strings---dividing by $4 \pi$ rather than $2\pi$---is motivated by the Euclidean thermodynamics. There, in the vicinity of the Misner string, the Euclidean time coordinate behaves like an angular coordinate on $\mathbb{S}^3$, and therefore has period $4\pi$ if regularity is enforced.}}
\be\label{psipm}
\psi_\pm=\frac{\kappa_\pm}{4\pi}=\frac{1}{8\pi (n\pm s)}\,.
\ee
Note that when $s=0$, the two coincide and we recover
\be\label{psisame}
\psi=\frac{1}{8\pi n}\,.
\ee

%%%%%%%%%%%%%%%%%%%%%%%%%%%%%%%%%%%%%%%%%%%%%%%%%%%%%%%%%%
%%%%%%%%%%%%%%%%%%%%%%%%%%%%%%%%%%%%%%%%%%%%%%%%%%%%%%%%%%
\section{Overview of thermodynamic argument for $s=0$}\label{sec3}
Before we proceed to identifying the Misner charges $N_\pm$ by Komar-type integration, let us briefly recapitulate the thermodynamic argument for formulating the thermodynamics of Taub--NUT solutions with $s=0$, as presented in  \cite{Kubiznak:2019yiu}.

The basic assumption in \cite{Kubiznak:2019yiu} was to identify the temperature of the spacetime with the temperature of the black hole horizon, \eqref{T}, and the entropy with the black hole horizon area
according to the Bekenstein--Hawking area law
\be
S=\frac{\mbox{Area}}{4}=\pi (r_+^2+n^2)\,.
\ee
The mass was calculated by the conformal method to yield
\be
M=m\,.
\ee

The next step was to calculate the Euclidean action \cite{Emparan:1999pm}:
\ba\label{action}
I&=&\frac{1}{16\pi}\int_{M}d^{4}x\sqrt{g}\bigl( R+\frac{6}{\ell^{2}}\bigr)\nonumber\\
&&+ \frac{1}{8\pi}\int_{\partial M}d^{3}x\sqrt{h}\left[\mathcal{K}
-   \frac{2}{\ell} - \frac{\ell}{2}\mathcal{R}\left( h\right) \right]\,,
\ea
and to associate it with a free energy, ${\cal G}=I/\beta$, $\beta=1/T$.
By assumption this was equated with
\be\label{Gco}
 {\cal G}={\cal G}(T, \psi, P)=M-TS-\psi N\,,
\ee
where $N$ was the new NUT-related (Misner) charge and $\psi$ the corresponding thermodynamic conjugate potential.
Specifically, the following result was  obtained in \cite{Kubiznak:2019yiu}  for the metric \eqref{j1}:
\be\label{G}
{\cal G}=\frac{I}{\beta}=\frac{m}{2}-\frac{r_+(r_+^2+3n^2)}{2l^2}\,.
\ee
We note here that the same expression remains valid also for the more general metric \eqref{metric2} with $s\neq 0$.

Using the explicit expression for ${\cal G}$, \eqref{G}, and the fact that all quantities apart from $\psi$ and $N$ on the r.h.s. of \eqref{Gco} are known, one can deduce that
\be
\psi N=-\frac{n^2}{2r_+}\Bigl(1+\frac{3(n^2-r_+^2)}{l^2}\Bigr)
\ee
Requiring the first law \eqref{first1} then fixes $\psi$ and $N$ uniquely (up to an overall reciprocal constant factor) and yields
\be\label{psiN}
\psi=\frac{1}{8\pi n}\,,\quad N=-\frac{4\pi n^3}{r_+}\Bigl(1+\frac{3(n^2-r_+^2)}{l^2}\Bigr)\,,
\ee
as well as the thermodynamic volume
\be\label{V}
V=\frac{4\pi r_+^3}{3}\Bigl(1+\frac{3n^2}{r_+^2}\Bigr)\,.
\ee

Obviously, the thermodynamic potential $\psi$ in \eqref{psiN} coincides with the potential $\psi$ introduced in \eqref{psisame}, given by the `surface gravity' of the symmetrically distributed Misner strings.
In what follows we generalize this notion and consider $\psi_\pm$ in \eqref{psipm} to be the new thermodynamic potentials associated with the Taub--NUT solutions. The remaining task is to find a geometric prescription for the conjugate quantities $N_\pm$. Since for general $s$ the potentials  $\psi_\pm$ are different so will be the conjugate quantities $N_\pm$. As we shall see, the new Misner charges $N_\pm$ can be obtained by the Komar-like integrals. In AdS some of the integrals have to be `renormalized'  \cite{Magnon:1985sc}. We do this here with the help of Killing co-potentials \cite{Kastor:2009wy}.

%%%%%%%%%%%%%%%%%%%%%%%%%%%%%%%%%%%%%%%%%%%
%%%%%%%%%%%%%%%%%%%%%%%%%%%%%%%%%%%%%%%%%%%
\section{Misner gravitational charges \& generalized Smarr}\label{sec4}
To find the Komar integral prescription for the Misner charges,
we re-derive the generalized Smarr relation for the Taub--NUT-AdS spacetime with arbitrary $s$ parameter, taking carefully into account the new
boundaries induced by the presence of Misner strings. For simplicity we concentrate on the static case in $d$ number of spacetime dimensions.

Consider a Killing vector $k=\partial_t$, a generator of the black hole horizon. Being a Killing vector, it satisfies the following two identities:
\be\label{ids}
\nabla_ak^a=0\,,\quad \nabla_a\nabla^ak^b=-R^b{}_ak^a\,.
\ee
The first one means \cite{Kastor:2009wy} that (at least locally) there exists a Killing co-potential 2-form $\omega$, such that
\be
k^a=\nabla_c \omega^{ca}\,.
\ee
Of course, such $\omega$ is not defined uniquely, and one can always perform
\be\label{freedom}
\omega\to \omega+\nu
\ee
where $\nabla_c \nu^{ca}=0$.
In an Einstein space,
\be
R_{ab}= \frac{2}{d-2} \Lambda g_{ab}\,,
\ee
the second identity \eqref{ids} then yields
\be
\nabla_a\nabla^ak^b=-\frac{2}{d-2} \Lambda k^b=-\frac{2}{d-2}\Lambda  \nabla_a \omega^{ab}\,,
\ee
 or in the language of differential forms
\begin{align}
d*dk+\frac{4}{d-2}\Lambda\ d*\omega=0\,.
\end{align}
Integrating over a $(d-1)$-dimensional hypersurface $\Sigma$ and using the Stokes' theorem, we thus recover
\be\label{integral}
0=\int_{\Sigma}d*\left(dk+\frac{4}{d-2}\Lambda\, \omega\right)
=\int_{\partial \Sigma}*\left(dk+\frac{4}{d-2}\Lambda\, \omega\right)\,.
\ee

\begin{figure}
\begin{center}
\includegraphics[width=0.42\textwidth,height=0.32\textheight]{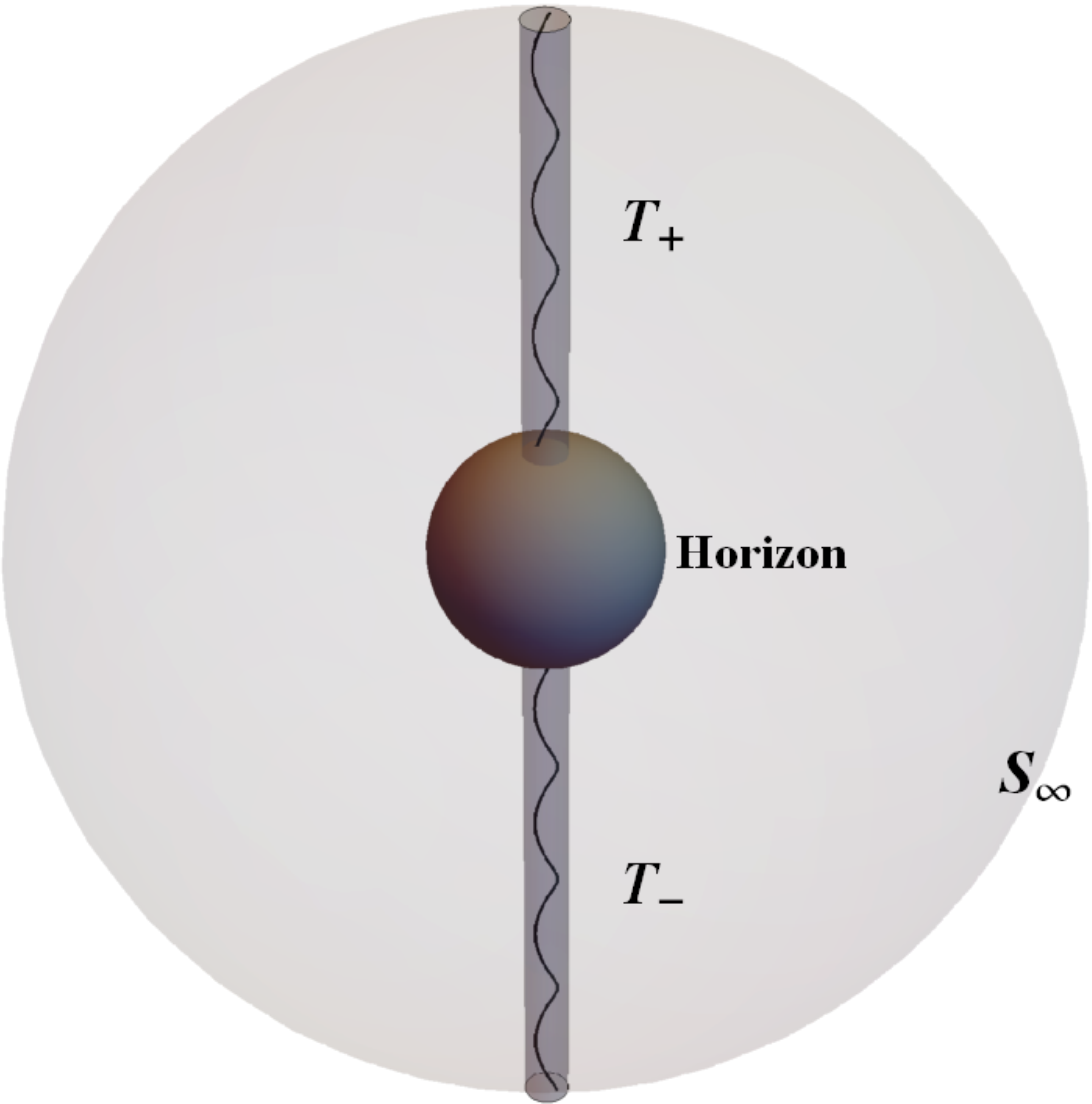}
\caption{{\bf Taub--NUT boundaries: Misner tubes.} In the Taub--NUT case, apart from the standard boundaries on the horizon $H$ and at infinity $S_\infty$, the two Misner tubes $T_\pm$ of radius $\epsilon$ around the strings are located at the north and south pole axis, $\cos\theta=\pm1$. The integrals over such Misner tubes define the Misner charges $N_\pm$ as well as modify the black hole volume $V$. For $s=\pm n$ the south/north pole axis becomes regular and the corresponding Misner tube no longer exists.
}\label{fig}
\end{center}
\end{figure}

For the vanishing NUT parameter this is a standard formula used for deriving the Smarr relation for AdS black hole spacetimes \cite{Kastor:2009wy}. Namely, choosing $\Sigma$ to be a $t=const$ hypersurface, the integral over $\partial\Sigma$ can be split into an integral over the horizon which gives $TS$ and $PV$ terms and the integral over a sphere at infinity which is related to the black hole mass $M$ \cite{Kastor:2009wy}.

In the case of the Taub--NUT, however, the integration becomes more subtle. Namely, when the Misner strings are present, $(\nabla t)^2$ is not well defined on the axis~\cite{misner1963flatter} and one is forced to introduce a new tube-like boundary around the Misner string singularities.
As illustrated in Fig.~\ref{fig}, the boundary $\partial \Sigma$ thus consists of the two Misner string tubes $T_+$ and $T_-$ located at $\theta=\epsilon$ and $\theta=\pi-\epsilon$ respectively, the black hole horizon $H$ at $r=r_+$, and the sphere $S_\infty$ at $r=\infty$. Taking into account the orientation of the boundaries we have
\be
\partial \Sigma= T_++S_{\infty}-T_--H\,.
\ee
The integral \eqref{integral} thus splits into the following five contributions:
\ba\label{eq:GemSmarr}
0&=&\int_{S_\infty}*\left(dk+\frac{4}{d-2}\Lambda \omega\right)-\int_{H}*dk \nonumber\\
&&{+}\Big(\int_{\tiny T_+} \!\!*dk+\frac{4}{d-2}\Lambda \Omega_+^{(r=\infty)}\Big)\nonumber\\
&&{-}\Big(\int_{\tiny T_-}\!\! *dk+\frac{4}{d-2}\Lambda \Omega_-^{(r=\infty)}\Big)\nonumber\\
&&-\frac{4}{d-2}\Lambda\Big(\int_{H}*\omega+\Omega_+^{(r=r_+)}-\Omega_-^{(r=r_+)}\Big)\,.\qquad
\ea
Here, the integration over the Misner tubes is over $r\in[r_+,\infty]$, $\phi\in[0,2\pi]$ (and other azimuthal angles in the case of $d>4$), and at fixed $t=const$. We have also introduced the notation
\be
\int_{\tiny T_\pm}*\omega=\Omega_\pm^{(r=\infty)}-\Omega_\pm^{(r=r_+)}\,,
\ee
splitting the integral into the $r\to \infty$ and $r=r_+$ parts.

The first two terms are standard and yield respectively the thermodynamic mass $M$ and the  product of entropy and temperature:
\ba
M&=& -\frac{(d-2)}{16\pi(d-3)}\int_{S_\infty}*\left(dk+\frac{4}{d-2}\Lambda\,\omega\right)\,,\label{Mint}\\
TS&=&-\frac{1}{16\pi}\int_{H}*dk\;.\label{TSint}
\ea

The arrangement of the other terms is novel. Namely, we define the gravitational Misner charges by
\be\label{Nint}
N_\pm\equiv\pm \frac{1}{16\pi \psi_\pm} \Big(\int_{\tiny T_\pm} \!\!*dk+\frac{4}{d-2}\Lambda \Omega_\pm^{(r=\infty)}\Big)\,.
\ee
Here, $\psi_\pm$ are the potentials given by the surface gravities of the corresponding Misner strings, the integral of Killing co-potential is only evaluated at $r=\infty$ and removes the divergence in the standard way of modified Komar integrals in AdS~\cite{Magnon:1985sc, Kastor:2009wy}.

The final integral in (\ref{eq:GemSmarr}) contains the Killing copotential evaluated on the horizon in the usual way plus the contributions from the Killing co-potential arising from the tube boundary evaluated at the horizon $r=r_+$. This integral yields the modified thermodynamic volume
\be\label{Vint}
V\equiv{-}\Big(\int_{H}*\omega+\Omega_+^{(r=r_+)}-\Omega_-^{(r=r_+)}\Big)\,.
\ee
With these definitions we thus derived the following generalized Smarr relation:
\be\label{Smarr}
(d-3)M=(d-2)(TS + \psi_+N_+ + \psi_-N_-)-2PV\;.
\ee

A word of warning is due here. As noted in \cite{Cvetic:2010jb} the freedom in the choice of the Killing co-potential, \eqref{freedom}, results in non-uniqueness of the prescription for mass $M$ and other quantities. ``The best one can do is to choose such $\omega$ so that the integral \eqref{Mint} yields the true thermodynamic mass.'' The ultimate check is the validity of the first law of thermodynamics \eqref{first2}.

Let us also note that integrals similar to \eqref{Nint}, renormailzed `by hand' instead of using the Killing co-potential, were considered in \cite{Garfinkle:2000ms}. In that case, however, they were interpreted as Misner string contributions to the total Noether charge entropy \cite{Garfinkle:2000ms} rather than independent Misner charges $N_\pm$.

%%%%%%%%%%%%%%%%%%%%%%%%%%%%%%%%%%%%%%%%%%%%%%%%%%%%%%%%%%
%%%%%%%%%%%%%%%%%%%%%%%%%%%%%%%%%%%%%%%%%%%%%%%%%%%%%%%%%%
\section{First law with variable Misner string strengths}\label{sec5}
Let us now illustrate the above definitions on the $(d=4$) Taub--NUT-AdS spacetime \eqref{metric2} with arbitrary distribution of Misner strings.
Such a metric admits a hidden symmetry encoded in the principal Killing--Yano tensor $h=db$, where \cite{Kubiznak:2007kh}
\be
b=-\frac{1}{2}r^2dt-{\bigl[n(r^2+n^2)\cos\theta+s r^2 \bigr]}d\phi\,.
\ee
Such a tensor underlies many remarkable properties of the spacetime \cite{Frolov:2017kze}. In particular, it gives the following Killing co-potential for the Killing vector $k=\partial_t$:
\ba\label{eq:KCP}
\omega=\frac{1}{3} h&=&
-\frac{r}{3} dr\wedge \bigl(dt+2[n\cos\theta+s]d\phi\bigr)\nonumber\\
&&+\frac{n}{3}(r^2+n^2)\sin\theta d\theta \wedge d\phi \,.
\ea
It can easily be shown that indeed $\nabla \cdot \omega =k$. Since $\sqrt{-g}=(r^2+n^2)\sin\theta$, the Hodge dual is
\ba
*\omega&=&-\frac{n}{3} dr\wedge \bigl(dt+2[n\cos\theta+s]d\phi\bigr)\nonumber\\
&&-\frac{r}{3}\sin\theta (r^2+n^2)d\theta \wedge d\phi\,.
\ea
At the same time we have
\ba
*dk&=&-\frac{2fn}{r^2+n^2}dr\wedge \bigl(dt+2[n\cos\theta+s]d\phi\bigr)\nonumber\\
&&-\sin\theta(r^2+n^2)f' d\theta\wedge d\phi\,.
\ea

It is now easy to show that the first integral \eqref{Mint} simply yields the mass $m$, while \eqref{TSint} yields
$TS=\frac{1}{4}f'(r_+)(r_+^2+n^2)$, all consistent with
\be
M=m\,,\quad T=\frac{f'(r_+)}{4\pi}\,,\quad S=\pi (r_+^2+n^2)\,.
\ee
The integrals \eqref{Nint} over the Misner tubes are a bit more interesting. We find
\ba
\int_{\tiny T_\pm} \!\!*dk&=&-\lim_{\epsilon\to 0}\int_{r=r_+}^\infty\frac{8\pi n(s\pm n\cos\epsilon)f}{r^2+n^2}dr\nonumber\\
&=&-8\pi n(s\pm n)\Bigl[\frac{r^3-3n^2r-rl^2+l^2m}{l^2(r^2+n^2)}\Bigr]^{r\to \infty}_{r=r_+}\,,\nonumber\\
\Omega_\pm^{(r=\infty)}&=&-\lim_{r\to \infty}\frac{4\pi n}{3}(s\pm n)r\,,
\ea
and the Killing co-potential part indeed renormalizes the first integral. Upon using $\psi_\pm$ in \eqref{psipm}, we thus find
\be
N_\pm=-\frac{2\pi n(n\pm s)^2}{r_+}\Bigl(1+\frac{3(n^2-r_+^2)}{l^2}\Bigr)\,.
\ee
Finally, considering the volume integral \eqref{Vint}, we find
\ba
\int_{H}*\omega&=&-\frac{4\pi r_+^3}{3}\Bigl(1+\frac{n^2}{r_+^2}\Bigr)\,,\nonumber\\
\Omega_+^{(r=r_+)}-\Omega_-^{(r=r_+)}&=&-\frac{8\pi n^2r_+}{3}\,,
\ea
and therefore
\be
V=\frac{4\pi r_+^3}{3}\Bigl(1+\frac{3n^2}{r_+^2}\Bigr)\,,
\ee
in accord with \eqref{V}.

It is then easy to verify that the obtained quantities satisfy the desired first law \eqref{first2},
\be\label{final}
\delta M=T\delta S+\psi_+\delta N_++\psi_-\delta N_-+V\delta P\,,
\ee
and by construction also the generalized Smarr relation \eqref{Smarr}. Note that the obtained first law
is of full cohomogeneity as all the parameters  $\{r_+,n,s, l\}$ may now be varied.
In other words, processes involving redistributing the Misner strings are allowed and described by the obtained first law.

Note also that for $s=0$ (symmetric distribution of Misner strings) the two $\psi_\pm$ coincide and yield the same $\psi$, \eqref{psisame}. In this case (and this case only) we can define new Misner charge
\be
N=N_++N_-=-\frac{4\pi n^3}{r+}\Bigl(1+\frac{3(n^2-r_+^2)}{l^2}\Bigr)\,,
\ee
and write the first law in the form \eqref{first1} presented in \cite{Kubiznak:2019yiu}.

Note also that for $s=+n$, the Misner string disappears on the south pole and \eqref{final} reduces to
\be\label{final2}
\delta M=T\delta S+\psi_+\delta N_++V\delta P\,,
\ee
and analogously for $s=-n$.

%%%%%%%%%%%%%%%%%%%%%%%%%%%%%%%%%%%%%%%%%%%%%%%%%
%%%%%%%%%%%%%%%%%%%%%%%%%%%%%%%%%%%%%%%%%%%%%%%%%%
\section{Some remarks on Noether charges}\label{sec6}
{The above derived extended first law \eqref{final} contains two new terms
\be
\psi_\pm \delta N_\pm\,,
\ee
where the Misner potentials $\psi_\pm$ are given by the surface gravity of the Killing horizons associated with the Misner strings and $N_\pm$ are the conjugate gravitational Misner charges.
It remains to be seen if (in Lorentzian signature) $\psi_\pm$ can be interpreted as a temperature. If this were the case, then the Lorentzian Taub-NUT solution would be somewhat analogous to de Sitter black holes, e.g. \cite{Dolan:2013ft}, constituting in general a multi-temperature system out of equilibrium. If such an interpretation were possible then the statement of the first law would be commensurate with statement that $N_+$ and $N_-$ are the entropies associated with the Misner strings.
}

{In the spirit of de Sitter analogy, one could then define a `total entropy'
\be
S_{\rm tot} = \mbox{Area}/4 + N_+ + N_-\,.
\ee
Note that this is different from what was in \cite{Kubiznak:2019yiu} called the `Noether charge entropy', $S_{\mbox{\tiny NC}}$, the difference between the two being given by the surface gravity factors:
\be
S_{\mbox{\tiny NC}}=\frac{\mbox{Area}}{4} +\frac{1}{T}(\psi_+N_++\psi_- N_-)\,.
\ee
Obviously, the two coincide only when the `equilibrium' is restored, imposing $\psi_+=\psi_-=T$, a situation studied in \cite{Garfinkle:2000ms}.
}

{Let us note, however, that provided $\psi_\pm$ truly correspond to temperatures, then it would seem more natural to define the entropy of each of the Misner horizons as Noether charge associated to $k_\pm$ (the generators of the horizons) normalized by the surface gravity factors, rather than $k=\partial_t$,
\be\label{Nint222}
\hat N_\pm\equiv\pm \frac{1}{16\pi \psi_\pm} \int_{\tiny T_\pm} \!\!*dk_\pm \,,
\ee
specializing for the moment to the asymptotically flat case where the Killing co-potentials do not appear.
This procedure is equivalent to the Lorentzian version of the argument in~\cite{Carlip:1999cy} and yields an entropy proportional to the (infinite) area of the strings:
\be
\hat N_\pm=\frac{\mbox{Area}_{\pm}}{2}=\lim_{R\to \infty} 2\pi (n\pm s)(R-r_+)\,,
\ee
where $R$ is the radial cutoff.
Using the fact that $k_\pm$, \eqref{kpm}, can be written as
\be
k_\pm=k\mp 4\pi \psi_\pm \eta\,,
\ee
(that is surface gravities of Misner strings are at the same time their angular velocities) yields
the following relation between $\hat N_\pm$ and $N_\pm$:
\be\label{formula222}
\hat N_\pm=N_\pm-\frac{1}{4}\int_{T^\pm}*d\eta\,.
\ee
It is somewhat interesting that the first law does not make use of this entropy, but rather only a ``portion'' of it encoded in $N_\pm$. Presumably this has to do with the fact that the first law is constructed for an observer whose asymptotic motion aligns with the generator of the black hole horizon, and not the generators of the string horizons. Note also that whereas $N_\pm$ {are finite their two contributions in \eqref{formula222} both diverge (unless `renormalized'  by for example the background substraction procedure).}
}

%%%%%%%%%%%%%%%%%%%%%%%%%%%%%%%%%%%%%%%%%%%%%%%%%%%
%%%%%%%%%%%%%%%%%%%%%%%%%%%%%%%%%%%%%%%%%%%%%%%%%%%%
\section{Summary and discussion}\label{sec7}
As shown recently, the thermodynamics of the Lorentzian Taub--NUT-AdS spacetimes with the Misner strings present can be consistently formulated \cite{Kubiznak:2019yiu}. The key ingredient is to introduce a new pair of conjugate quantities $\psi-N$. However, in \cite{Kubiznak:2019yiu} no physical intuition or geometrical interpretation for these quantities was offered and the entire construction based on thermodynamic considerations, namely that the full cohomogeneity first law where the NUT parameter is independently varied should be valid.

In this paper we have extended these results in a number of ways.
In particular, we have identified the new potential $\psi$ with the surface gravity of the Misner string Killing horizon that is present in the spacetime. For a symmetric distribution of Misner strings (the case of $s=0$ studied in \cite{Kubiznak:2019yiu}) such a surface gravity is the same on the north and south pole axes. However, for a general Taub--NUT spacetime the strengths of the two Misner strings will be different and so will be their corresponding surface gravities, leading thus to two independent potentials $\psi_\pm$. We have shown that the corresponding thermodynamically conjugate charges $N_\pm$ can then be obtained by a Komar-type integration over the tubes surrounding the Misner string singularities \eqref{Nint}. This establishes the geometrical meaning of $N_\pm$ {(though the physical meaning, for example an interpretation in terms of entropy, remains to be seen)}.
Similar tube contributions also modify the `standard formula' \cite{Kastor:2009wy, Cvetic:2010jb} for the geometric black hole volume $V$, see \eqref{Vint}.
To renormalize the necessary divergences present in the AdS case we have used the method of Killing co-potentials \cite{Kastor:2009wy}. The resultant charges are then automatically finite and by construction satisfy the generalized Smarr relation \eqref{Smarr}.

We have also extended the first law \eqref{first1} presented in \cite{Kubiznak:2019yiu} to a more general law \eqref{first2}, allowing for the asymmetric distribution of Misner strings. In fact, the parameter $s$ (governing this distribution) can now be freely varied in \eqref{first2}. Such a law thus allows for variable Misner string strengths and in that sense is of higher cohomogeneity than \eqref{first1}. The variable string strengths can perhaps be realized in a process of capture of a Misner string by a black hole or an idealized axisymmetric merger of two Taub--NUT solutions.

The fact that we have explicitly shown that the thermodynamic volume receives contributions from the Misner strings raises further interesting questions concerning the reverse isoperimetric inequality~\cite{Cvetic:2010jb}.\footnote{Recall that it was conjectured in~\cite{Cvetic:2010jb} that for a black hole with area ${\cal A}$ and thermodynamic volume ${\cal V}$ the ratio \[{\cal R} = \left(\frac{(d-1) {\cal V}}{{\cal A}_{d-2}}\right)^{1/(d-1)}\left(\frac{{\cal A}_{d-2}}{{\cal A}}\right)^{1/(d-2)}, \] satisfies ${\cal R} \ge 1$.  } Namely, it is natural to wonder whether the `area' that appears in this formula is simply the area of the black hole horizon while the volume is the total thermodynamic volume of the spacetime, or whether the area should receive additional contributions from the Misner strings. Yet another possibility is that the horizon and Misner strings should be considered separately in the context of this inequality, similar to the event and cosmological horizons for de Sitter black holes~\cite{Dolan:2013ft}.  If this is the case, the reverse isoperimetric inequality would continue to hold for the horizon, while, since the `area' of the Misner strings is infinite, it would seem to imply that it is the \textit{isoperimetric} inequality that holds for them, analogous to the volume between the black hole and cosmological horizons in the de Sitter case.

Let us finally mention that the thermodynamic charges $N_\pm$ as well as the volume $V$ can  also be obtained by pure thermodynamic considerations, from the
Euclidean action \eqref{action}, once the potentials $\psi_\pm$ have been identified. Namely, as noted in Sec.~\ref{sec3}, for general $s$ the free energy ${\cal G}$ corresponding to the Euclidean action \eqref{action} remains given by \eqref{G}.
Using \eqref{psipm} for $\psi_\pm$, \eqref{T} for $T$, and \eqref{P} for $P$, one can explicitly write
\be
{\cal G}={\cal G}(T,\psi_+,\psi_-, P)\,.
\ee
The conjugate thermodynamic quantities are then given by
\be\label{jinak}
S=-\frac{\partial {\cal G}}{\partial T}\,,\quad N_\pm=-\frac{\partial {\cal G}}{\partial \psi_\pm}\,,\quad V=\frac{\partial {\cal G}}{\partial P}\,.
\ee
By construction these satisfy the first law \eqref{first2}. The Smarr relation \eqref{Smarr} is then equivalent to the relation ${\cal G}=M-TS-\psi_+N_+-\psi_-N_-$. Of course, one can easily check that the thermodynamic quantities \eqref{jinak} coincide with the geometric quantities derived in the previous section.   {This approach is similar to that used in~\cite{Bueno:2018uoy} where it was found that, in the absence of Misner strings, such terms are necessary for consistent thermodynamics of Taub solutions with toroidal base spaces.}

%\newpage

\section*{Acknowledgements}
We are grateful to Robert Mann for
discussions.
This work was supported in part by the Natural Sciences and Engineering Research Council of Canada.
R.H. is grateful to the Banting
Postdoctoral Fellowship programme.
A.B., F.G., and D.K.\ acknowledge the Perimeter Institute for Theoretical Physics  for their support. Research at Perimeter Institute is supported by the Government of Canada through the Department of Innovation, Science and Economic Development Canada and by the Province of Ontario through the Ministry of Research, Innovation and Science.

\appendix

%%%%%%%%%%%%%%%%%%%%%%%%%%%%%%%%%%%%%%%%%%%%%%%%%%%%%%%%%%%
%%%%%%%%%%%%%%%%%%%%%%%%%%%%%%%%%%%%%%%%%%%%%%%%%%%%%%%%%%%%%%
\section{Generalized first law for the dyonic Taub--NUTs}
In this appendix we extend the recent results on thermodynamics of dyonic Taub--NUTs of the Einstein--Maxwell theory \cite{Ballon:2019uha}, to account for asymmetric distributions of Misner strings and their potential variable strengths. To this purpose we shall not generalize the Komar integration to the presence of matter but rather take a short cut and use the Euclidean action.
%\tcr{\bf Could someone check all the formulas in this section please?}

The dyonic Taub--NUT--AdS solution with arbitrary parameter $s$ reads
\ba\label{j1}
ds^2&=&-f\bigl[dt+2(n\cos\theta +s) d\phi\bigr]^2+\frac{dr^2}{f}\nonumber\\
&&\qquad \quad +(r^2+n^2)(d\theta^2+\sin^2\!\theta d\phi^2)\,,\nonumber\\
A&=&-h\bigl[dt+2(n\cos\theta+s) d\phi\bigr]\,,
\ea
with
\ba\label{j2}
f&=&\frac{r^2-2mr-n^2+4n^2g^2+e^2}{r^2+n^2}-\frac{3n^4-6n^2r^2-r^4}{l^2(r^2+n^2)}\,,\nonumber\\
h&=&\frac{er}{r^2+n^2}+\frac{g(r^2-n^2)}{r^2+n^2}\,,
\ea
where $e$ and $g$ are the electric and magnetic parameters, while other parameters assume the meaning as before.

The black hole horizon, located at $f(r_+)=0$, is generated by $k=\partial_t$ and has the following temperature and entropy
\ba
T&=&\frac{f'}{4\pi}=\frac{1}{4\pi r_+}\Bigl(1+\frac{3(r_+^2+n^2)}{l^2}-\frac{e^2+4n^2g^2}{r_+^2+n^2}\Bigr)\,,\label{T33}\quad\\
S&=&\frac{\mbox{Area}}{4}=\pi (r_+^2+n^2)\,.\label{S33}
\ea
As in the uncharged case, the Misner string horizons are generated by $\xi_\pm=\partial_t-\frac{1}{2(n\pm s)}\partial_\phi$\,, and give rise to the following
Misner potentials:
\be\label{eq:psi}
\psi_\pm=\frac{\kappa_\pm}{4\pi}=\frac{1}{8\pi(n\pm s)}\;.
\ee

The parameters $e$ and $g$ are related to the electric $q$ and magnetic $q_m$ charges by the Gauss's law:
%. Namely, employing the Gauss's law, the electric charge $q$ and the magnetic charge $q_m$ are given by
\be
q=\frac{1}{4\pi} \int_{S^2} *F\,,\quad q_m=\frac{1}{4\pi}\int_{S^2} F\,,
\ee
where $F=dA$ is the Maxwell field strength.
By performing the integration over a sphere of radius $r$, we find
\be\label{qqm}
q=\frac{e(r^2-n^2)-4gr n^2}{r^2+n^2}\,,\quad q_m=-\frac{2n\bigl(er+g[r^2-n^2]\bigr)}{r^2+n^2}\,,
\ee
which, contrary to common wisdom and known examples of dyonic solutions depends on the radius of the sphere.
Remarkably, as found in \cite{Ballon:2019uha} these charges enter the first law in a way that one of these is evaluated at infinity and the other on the horizon. For the purpose of this appendix we take
\ba
Q&\equiv& q(r\to\infty)=e\,,\label{Qe}\\
Q_m^{(+)}&\equiv& q_m(r=r_+)=-\frac{2n\bigl(er_++g[r_+^2-n^2]\bigr)}{r_+^2+n^2}\,.\label{Qm}
\ea
The electrostatic potential is
\be\label{phi}
\phi=-\bigl(k\cdot A|_{r=r_+}-k\cdot A|_{r=\infty}\bigr)=\frac{er_+-2gn^2}{r_+^2+n^2}\,.
\ee
Employing the `electromagnetic duality'
\be\label{EM}
e\leftrightarrow -2ng\,,\quad 2ng\leftrightarrow e\,,
\ee
which relates $q\leftrightarrow q_m$, the corresponding magnetic potential is
\be\label{phim}
\phi_m=-\frac{n(2gr_++e)}{r_+^2+n^2}\,.
\ee
%for the thermodynamic magnetostatic potential.

Turning now towards the action
\ba
I&=&\frac{1}{16\pi}\int_{M}d^{4}x\sqrt{g}\left[ R+\frac{6}{ l^{2}}-F^2\right]
\nonumber\\
&&+ \frac{1}{8\pi}\int_{\partial M}d^{3}x\sqrt{h}\left[\mathcal{K}
-   \frac{2}{ l} - \frac{ l}{2}\mathcal{R}\left( h\right) \right]\,,
\label{action}
\ea
one finds
\ba\label{Free}
{\cal G}&=&\frac{m}{2}-\frac{r_+(3n^2+r_+^2)}{2l^2}-\frac{r_+e^2(r_+^2-n^2)}{2(n^2+r_+^2)^2}\nonumber\\
&&+\frac{2n^2r_+(r_+^2-n^2)g^2}{(r_+^2+n^2)^2}+\frac{4n^2 eg r_+^2}{(n^2+r_+^2)^2}\,.
\ea
The action as written corresponds to the grand-canonical (fixed electric potential and fixed magnetic charge) ensemble and so is a function of the following quantities: %cite{Caldarelli:1999xj} and so is a function of these quantities as well as the temperature and ``Misner potentials''
\be
{\cal G}={\cal G}(T,\psi_+,\psi_-,\phi,Q_m^{(+)},P)\;.
\ee
The corresponding thermodynamic quantities are then given by
\begin{align}
S&=-\frac{\partial \mathcal{G}}{\partial T}\,,\quad N_+=-\frac{\partial \mathcal{G}}{\partial \psi_+}\,,\quad N_-=-\frac{\partial \mathcal{G}}{\partial \psi_-}\;,\nonumber\\
Q&=-\frac{\partial \mathcal{G}}{\partial \phi}\quad \phi_m=\frac{\partial \mathcal{G}}{\partial Q_m^{(+)}}\,,\quad V=\frac{\partial \mathcal{G}}{\partial P}\,.
\end{align}.
This yields $S$ in \eqref{S33}, $Q$ in \eqref{Qe}, $\phi_m$ in \eqref{phim}, and $V$ in \eqref{V}. We also find the following Misner charges:
\ba
N_\pm&=&-\frac{4\pi n(n\pm s)^2}{r_+}\Bigl(1+\frac{3(n^2-r_+^2)}{l^2}\nonumber\\ &&+\frac{(r_+^2-n^2)(e^2+4egr_+)}{(r_+^2+n^2)^2}-\frac{4n^2g^2(3r_+^2+n^2)}{(r_+^2+n^2)^2}\Bigr)\,.\nonumber\\
\ea

Identifying $M=m$, this yields the following full cohomogeneity first law:
\ba
\delta M&=&T\delta S+\phi \delta Q+\phi_m \delta Q_m^{(+)}\nonumber\\
&&\;\;\;\;\;\;\;\;\quad + \psi_+ dN_+ + \psi_- dN_- + V \delta P\,,\label{first}\quad
\ea
where all 6 parameters of the solution: $\{n,m, e, g, s, l\}$ are independently varied.
The relation
${\cal G}=M-TS-\phi Q -\psi_+ N_+-\psi_-N_-$ is then equivalent to the Smarr relation
\be
M=2(TS-VP+\psi_+ N_+ +\psi_- N_- )+\phi Q+\phi_m Q_m^{(+)}\,.\label{Smarr2}
\ee
The first law with $Q\to Q^{(+)}$ and $Q_m^{(+)}\to Q_m$ could be obtained analogously and so could be the special (constrained) subcases studied in \cite{Ballon:2019uha}.

\providecommand{\href}[2]{#2}\begingroup\raggedright\endgroup


\begin{thebibliography}{10}

\bibitem{Kubiznak:2019yiu}
R.~A. Hennigar, D.~Kubiznak, and R.~B. Mann, {\it {Thermodynamics of Lorentzian
  Taub-NUT spacetimes}},  \href{http://xxx.lanl.gov/abs/1903.0866}{{\tt
  arXiv:1903.0866}}.

\bibitem{taub1951empty}
A.~H. Taub, {\it Empty space-times admitting a three parameter group of
  motions},  {\em Annals of Mathematics} (1951) 472--490.

\bibitem{newman1963empty}
E.~Newman, L.~Tamburino, and T.~Unti, {\it Empty-space generalization of the
  Schwarzschild metric},  {\em Journal of Mathematical Physics} {\bf 4} (1963),
  no.~7 915--923.

\bibitem{misner1963flatter}
C.~W. Misner, {\it The flatter regions of Newman, Unti, and Tamburino's
  generalized Schwarzschild space},  {\em Journal of Mathematical Physics} {\bf
  4} (1963), no.~7 924--937.


\bibitem{bonnor1969new}
W.~B. Bonnor, {\it A new interpretation of the NUT metric in general
  relativity},  in {\em Mathematical Proceedings of the Cambridge Philosophical
  Society}, vol.~66, pp.~145--151, Cambridge University Press, 1969.

\bibitem{Israel:1976vc}
W.~Israel, {\it {Line sources in general relativity}},  {\em Phys. Rev.} {\bf
  D15} (1977) 935--941.

\bibitem{Clement:2015cxa}
G.~Cl\'{e}ment, D.~Gal'tsov, and M.~Guenouche, {\it {Rehabilitating space-times
  with NUTs}},  {\em Phys. Lett.} {\bf B750} (2015) 591--594,
  [\href{http://xxx.lanl.gov/abs/1508.0762}{{\tt arXiv:1508.0762}}].

\bibitem{Clement:2015aka}
G.~Cl\'{e}ment, D.~Gal'tsov, and M.~Guenouche, {\it {NUT wormholes}},  {\em Phys.
  Rev.} {\bf D93} (2016), no.~2 024048,
  [\href{http://xxx.lanl.gov/abs/1509.0785}{{\tt arXiv:1509.0785}}]. [Phys.
  Rev.D93,024048(2016)].

\bibitem{Clement:2016mll}
G.~Cl\'{e}ment and M.~Guenouche, {\it {Motion of charged particles in a NUTty
  Einstein-Maxwell spacetime and causality violation}},  {\em Gen. Rel. Grav.}
  {\bf 50} (2018), no.~6 60, [\href{http://xxx.lanl.gov/abs/1606.0845}{{\tt
  arXiv:1606.0845}}].

\bibitem{miller1971taub}
J.~Miller, M.~D. Kruskal, and B.~B. Godfrey, {\it Taub-NUT (Newman, Unti,
  Tamburino) metric and incompatible extensions},  {\em Physical Review D} {\bf
  4} (1971), no.~10 2945.

\bibitem{Ballon:2019uha}
A.~B. Bordo, F.~Gray, and D.~Kubiznak, {\it {Thermodynamics and Phase
  Transitions of NUTty Dyons}},  \href{http://xxx.lanl.gov/abs/1904.0003}{{\tt
  arXiv:1904.0003}}.

\bibitem{Garfinkle:2000ms}
D.~Garfinkle and R.~B. Mann, {\it {Generalized entropy and Noether charge}},
  {\em Class. Quant. Grav.} {\bf 17} (2000) 3317--3324,
  [\href{http://xxx.lanl.gov/abs/gr-qc/0004056}{{\tt gr-qc/0004056}}].

\bibitem{Carlip:1999cy}
S.~Carlip, {\it {Entropy from conformal field theory at Killing horizons}},
  {\em Class. Quant. Grav.} {\bf 16} (1999) 3327--3348,
  [\href{http://xxx.lanl.gov/abs/gr-qc/9906126}{{\tt gr-qc/9906126}}].

\bibitem{Kastor:2009wy}
D.~Kastor, S.~Ray, and J.~Traschen, {\it {Enthalpy and the Mechanics of AdS
  Black Holes}},  {\em Class. Quant. Grav.} {\bf 26} (2009) 195011,
  [\href{http://xxx.lanl.gov/abs/0904.2765}{{\tt arXiv:0904.2765}}].

\bibitem{Anabalon:2018qfv}
  A.~Anabalon, F.~Gray, R.~Gregory, D.~Kubiznak and R.~B.~Mann,
  {\it Thermodynamics of Charged, Rotating, and Accelerating Black Holes},
  {\em JHEP} {\bf 1904} (2019) 096,
  [\href{http://xxx.lanl.gov/abs/1811.04936}{{\tt arXiv:1811.04936}}].


\bibitem{Emparan:1999pm}
R.~Emparan, C.~V. Johnson, and R.~C. Myers, {\it {Surface terms as counterterms
  in the AdS / CFT correspondence}},  {\em Phys. Rev.} {\bf D60} (1999) 104001,
  [\href{http://xxx.lanl.gov/abs/hep-th/9903238}{{\tt hep-th/9903238}}].

\bibitem{Magnon:1985sc}
A.~Magnon, {\it {On Komar integrals in asymptotically anti-de Sitter
  space-times}},  {\em J. Math. Phys.} {\bf 26} (1985) 3112--3117.

\bibitem{Cvetic:2010jb}
M.~Cvetic, G.~W. Gibbons, D.~Kubiznak, and C.~N. Pope, {\it {Black Hole
  Enthalpy and an Entropy Inequality for the Thermodynamic Volume}},  {\em
  Phys. Rev.} {\bf D84} (2011) 024037,
  [\href{http://xxx.lanl.gov/abs/1012.2888}{{\tt arXiv:1012.2888}}].

\bibitem{Kubiznak:2007kh}
D.~Kubiznak and P.~Krtous, {\it {On conformal Killing-Yano tensors for
  Plebanski-Demianski family of solutions}},  {\em Phys. Rev.} {\bf D76} (2007)
  084036, [\href{http://xxx.lanl.gov/abs/0707.0409}{{\tt arXiv:0707.0409}}].

\bibitem{Frolov:2017kze}
V.~Frolov, P.~Krtous, and D.~Kubiznak, {\it {Black holes, hidden symmetries,
  and complete integrability}},  {\em Living Rev. Rel.} {\bf 20} (2017), no.~1
  6, [\href{http://xxx.lanl.gov/abs/1705.0548}{{\tt arXiv:1705.0548}}].

\bibitem{Dolan:2013ft}
B.~P. Dolan, D.~Kastor, D.~Kubiznak, R.~B. Mann, and J.~Traschen, {\it
  {Thermodynamic Volumes and Isoperimetric Inequalities for de Sitter Black
  Holes}},  {\em Phys. Rev.} {\bf D87} (2013), no.~10 104017,
  [\href{http://xxx.lanl.gov/abs/1301.5926}{{\tt arXiv:1301.5926}}].

\bibitem{Bueno:2018uoy}
P.~Bueno, P.~A. Cano, R.~A. Hennigar, and R.~B. Mann, {\it {NUTs and bolts
  beyond Lovelock}},  {\em JHEP} {\bf 10} (2018) 095,
  [\href{http://xxx.lanl.gov/abs/1808.0167}{{\tt arXiv:1808.0167}}].

\end{thebibliography}
\end{document}